\documentclass[dvips]{article}

\usepackage{icrc2011}
\usepackage{amsmath,amssymb}
\usepackage{url}

\title{Quantifying uncertainties in the high energy neutrino cross-section}

\newcommand{\etal}{\MakeLowercase{\textit{et al. }}} 
\shorttitle{Cooper-Sarkar \etal Quantifying uncertainties in the high energy neutrino cross-section}

\authors{Amanda Cooper-Sarkar$^{1}$, Philipp Mertsch$^{2}$, Subir
  Sarkar$^{2}$}
\afiliations{$^1$Particle Physics, University of Oxford, Keble Road,
  Oxford OX1 3RQ, UK\\$^2$Rudolf Peierls Centre for Theoretical
  Physics, University of Oxford, 1 Keble Road, Oxford OX1 3NP, UK\\}
\email{p.mertsch1@physics.ox.ac.uk}

\abstract{We compare predictions for high energy neutrino and
  anti-neutrino deep inelastic scattering cross-sections within the
  conventional DGLAP formalism of next-to-leading order QCD, using the
  latest parton distribution functions such as CT10, HERAPDF1.5 and
  MSTW08 and taking account of PDF uncertainties. From this we derive
  a benchmark cross-section and uncertainty which is consistent with
  the results obtained earlier using the ZEUS-S PDFs. We advocate the use of this for analysing data from neutrino telescopes, in order to facilitate comparison between their results.}

\begin{document}
\maketitle

\section{Introduction}

Searches for high-energy cosmic neutrinos rely on predictions for
  the neutrino cross-section at high energies. These have however
  sizeable uncertainties deriving from the uncertainties on the parton
  distribution functions (PDFs) of the nucleon. Conventional PDF fits
use the next-to-leading-order (NLO) DGLAP
formalism~\cite{Altarelli:1977zs,Gribov:1972ri,Lipatov:1974qm,Dokshitzer:1977sg}
of QCD to make predictions for DIS cross-sections of leptons on
hadrons. At low $x$ it may be necessary to go beyond the DGLAP
formalism in order to sum $\ln(1/x)$ diagrams, as in the BFKL
formalism~\cite{Kuraev:1977fs,Balitsky:1978ic,Lipatov:1985uk} (for
recent work see
Refs.~\cite{Rojo:2009us,Caola:2009iy}), or to even consider non-linear
terms as in the colour glass condensate
model~\cite{Gelis:2010nm,Goncalves:2010ay}. While the exact
  theoretical framework at low $x$ is still contested it has been
  suggested~\cite{Anchordoqui:2006ta,Kusenko:2001gj,Anchordoqui:2005ey,PalomaresRuiz:2005xw,Anchordoqui:2010hq}
  that observations of ultra high-energy neutrinos could itself be
  used to measure the cross-section thereby constraining the
  models. It is therefore important to not only consider the
  prediction for the cross-section but also to estimate their
  uncertainties in the conventional NLO DGLAP formalism.

In the framework of the quark-parton model, high energy neutrino deep
inelastic scattering (DIS) accesses large values of $Q^2$, the
invariant mass of the exchanged vector boson, and small values of
Bjorken $x$, the fraction of the momentum of the incoming nucleon
taken by the struck quark. Thus in evaluating uncertainties on high
energy neutrino DIS cross-sections it is important to use the most
up-to-date information from the experiments at HERA, which have
accessed the lowest $x$ and highest $Q^2$ scales to date. H1 and ZEUS
have now combined the data collected in the years 1994--2000 to give
very accurate inclusive cross-sections in the range $ 6 \times 10^{-7}
< x < 0.65 $ and $ 0.045 < Q^2 < 30000$
GeV$^2$~\cite{Aaron:2009wt}. These data have not been available
  (or not been used) in previous
  predictions~\cite{Gandhi:1998ri,CooperSarkar:2007cv,Connolly:2011vc}.
It is the purpose of the present paper to re-evaluate the high energy
cross-sections using the most up-to-date PDF sets, with particular
emphasis on those which do use these precise, combined HERA data. The
calculation is made using PDFs which were evaluated in NLO DGLAP fits,
and our calculation of the neutrino structure functions and
cross-sections is also made consistently at NLO. For further details,
we refer the interested reader to Ref.~\cite{CooperSarkar:2011pa}.

\section{Formalism}

The kinematics of lepton hadron scattering is described in terms of
the variables $Q^2$, Bjorken $x$, and $y$ which measures the energy
transfer between the lepton and hadron systems.  The double
differential charged current (CC) cross-section for neutrino and
anti-neutrino production on isoscalar nucleon targets is given by
\cite{Devenish:2004pb}
\begin{equation*}
 \frac{\mathrm{d}^2\sigma}{\mathrm{d}x~\mathrm{d}Q^2} = 
 \frac{G_\mathrm{F}^2 M_W^4}{4\pi(Q^2 + M_W^2)^2 x} 
 \sigma_\mathrm{r},
\label{eqn:sigma}
\end{equation*}
where the reduced cross-section $\sigma_\mathrm{r}(\nu (\bar{\nu})
N)$ is 
\begin{equation*} 
 \sigma_\mathrm{r} = 
 \left[Y_ + F_2^{\nu} (x, Q^2) - y^2 F_\mathrm{L}^{\nu} (x, Q^2) 
 + Y_ - xF_3^{\nu} (x, Q^2) \right],
\label{eqn:nu}
\end{equation*}
and $F_2$, $xF_3$ and $F_\mathrm{L}$ are related directly to quark
momentum distributions, with $Y_{\pm} = 1 \pm (1-y)^2$.

The QCD predictions for these structure functions are obtained by
solving the DGLAP evolution equations at NLO in the
\mbox{$\overline{\mathrm{MS}}$} scheme with the renormalisation and
factorization scales both chosen to be $Q^2$.  These equations yield
the PDFs at all values of $Q^2$ provided these distributions have been
input as functions of $x$ at some input scale $Q^2_0$.

In QCD at leading order, the structure function $F_\mathrm{L}$ is
identically zero, and the structure functions $F_2$ and $xF_3$ for
charged current neutrino interactions on isoscalar targets can be
identified with quark distributions.  At NLO these expressions must be
convoluted with appropriate co-efficient functions in order to obtain
the structure functions (and $F_\mathrm{L}$ is no longer zero) but
these expressions still give us a good idea of the dominant
contributions. Cross-sections for neutral current (NC) and anti-neutrino interactions are calculated in a similar way.

\section{Parton Density Functions}

Uncertainties on PDFs derive from two sources: experimental errors
  and parametrisation uncertainties. To allow for the estimation of
  the error induced in the predicted observable, i.e. cross-sections
  in the present case, modern PDF sets provide not only the best-fit
  PDF but also variants that reflect these different
  uncertainties. For experimental errors a set of variant PDFs,
  so-called eigenvectors, is obtained after diagonalisation of the
  error matrix. The eigenvectors are linearly independent such that
  the individual experimental errors can be added in quadrature. The
  variants for the parametrisation uncertainties are obtained from
  fits by varying certain parameter values (e.g. the starting scale
  $Q^2_0$ for evolution and the value of $\alpha_\mathrm{s} (M_Z)$) or
  the parametrisation for the input PDF parametrisation at $Q^2_0$.

The PDF4LHC group has recently benchmarked modern parton density
functions~\cite{Alekhin:2011sk}.  Since our concern is with high
energy neutrino cross-sections, rather than with LHC physics, we focus
on PDF sets which make use of the newly combined accurate HERA data
\cite{Aaron:2009wt}. Of all the PDFs considered by the PDF4LHC only
HERAPDF1.0 \cite{Aaron:2009wt}) and NNPDF2.0 \cite{Ball:2010de} used
these data. However there has been a subsequent update of the CTEQ6.6
\cite{Tung:2006tb} PDFs to CT10~\cite{Lai:2010vv} which does use these
data, while HERAPDF1.0 has recently updated to
HERAPDF1.5~\cite{CooperSarkar:2010wm} using an preliminary combination
of HERA data from 2003--2007 as well as the published combined
data. We will utilise the CT10 and HERAPDF1.5 PDFs for the present
study; we also consider the MSTW2008 PDFs in order to compare with
other recent calculations of high energy neutrino
cross-sections~\cite{Connolly:2011vc}, although we caution that these
have \emph{not} included the most accurate HERA low $x$ data relevant
to the present study.

\section{Results}

The calculation of the CC and NC cross-sections in NLO has been
performed using \texttt{DISPred}~\cite{Ferrando:2010dx}. The PDFs
  have been implemented through the LHAPDF
  interface~\cite{LHAPDF}. Particular care has been exercised to
  perform a self-consistent calculation. For example the PDFs from
  LHAPDF are mostly defined for a limited range in $Q^2$ and $x$ and
  ``freeze'' beyond this range, which would result in underestimation of the cross-section at
  high energies; therefore we have used other
  implementations~\cite{CTEQ,MSTW}. Naturally, the cross sections have
  been calculated at a consistent order with respect to the PDFs.

\begin{figure}[!t]
  \vspace{5mm}
  \centering
  \includegraphics[width=\columnwidth]{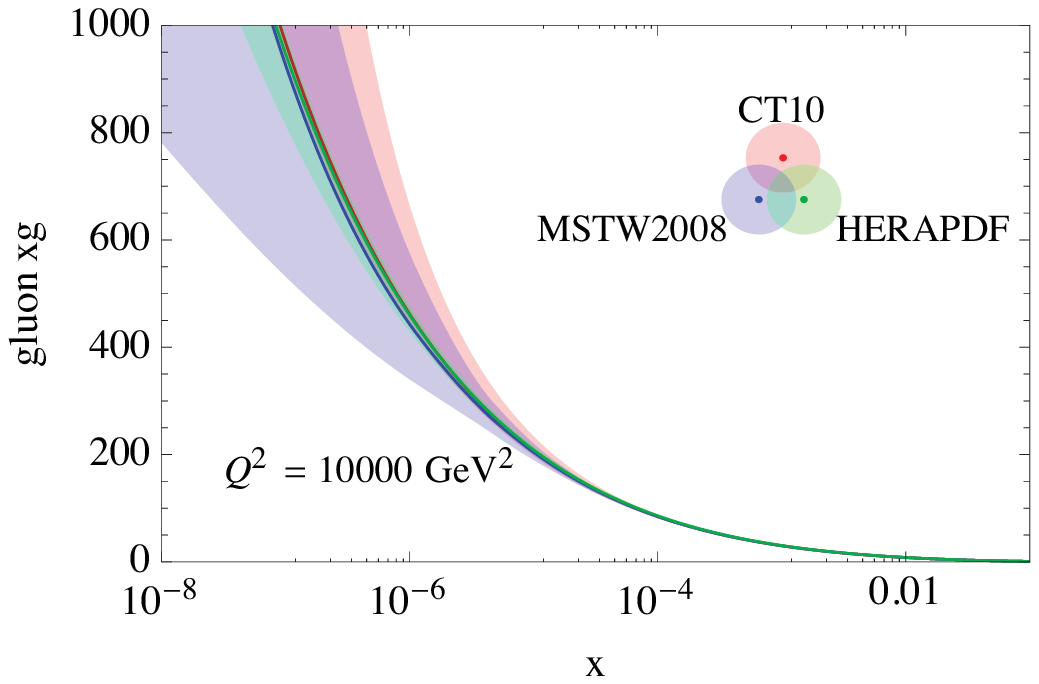}
  \includegraphics[width=\columnwidth]{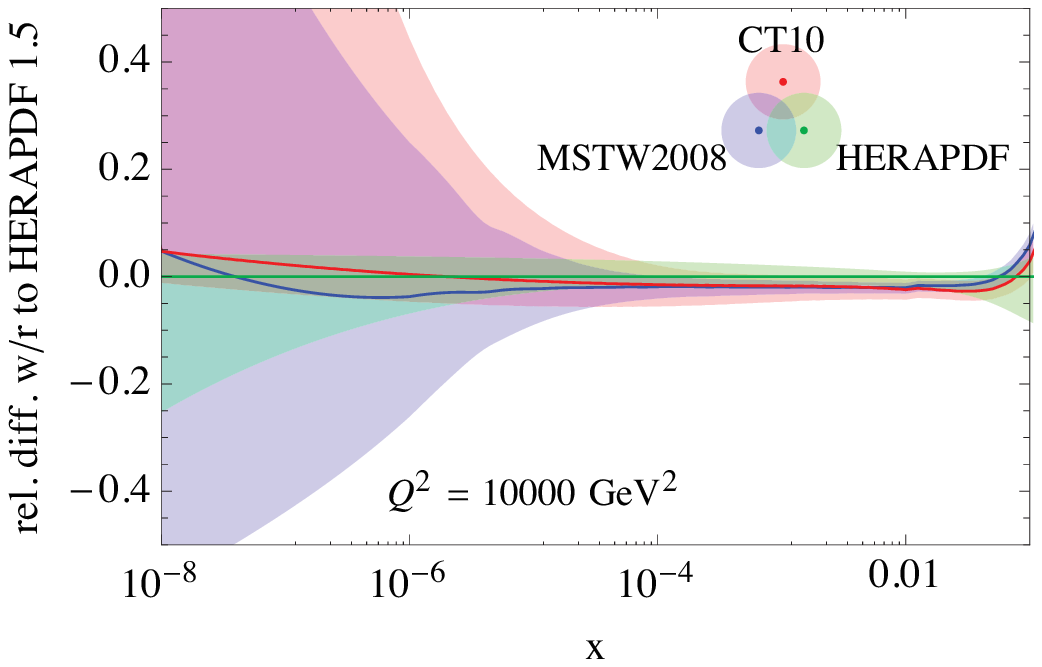}
  \caption{{\bf Top panel:} Gluon structure function at $Q^2 =
    10^4\,\text{GeV}^2$ for the three PDF sets used. {\bf Bottom
      panel:} The relative deviations and uncertainties (at 68\% c.l.)
    with respect to the central value of HERAPDF1.5. The uncertainty
    bands are shown \emph{with} member 9 for HERAPDF1.5 and member 52
    for CT10.}
  \label{fig:glucomp}
 \end{figure}

Figure~\ref{fig:glucomp} compares the gluon PDF and its uncertainty at
$Q^2=10000$~GeV$^2$ for the three PDFs which we consider. This value
of $Q^2$ is in the middle of the range which contributes significantly
to the neutrino cross-sections. We see that the central values of the
gluon PDFs are all very similar, whereas the uncertainty estimates
differ.  The CT10 and HERAPDF1.5 uncertainties are actually very
similar if we leave out member 52 from the CT10 error set. This error
set was introduced into the CT10 analysis to allow for a larger
uncertainty at low $x$~\cite{P.Nadolskyprivatecomm}. Previous CTEQ
analyses such as CTEQ6.6 \cite{Tung:2006tb} do not have such an
extreme error set.  The problem with such an \emph{ad hoc}
introduction of a steeply increasing gluon PDF is that at low $x$ it
leads to a very strong rise of the 
unphysical.

The larger error band of MSTW2008 is partly due to the fact that it
does not include the most up to date HERA data, which have
significantly reduced errors at low $x$. However the more striking
difference between MSTW2008 and both HERAPDF1.5 and CT10 is the
downward divergence of its error band which is due to the gluon
becoming negative at low $x,\,Q^2$. At NLO the gluon PDF does not have
to be positive, although one might consider that it going negative
signals a breakdown of the DGLAP formalism.  However measurable
quantities such as the longitudinal structure function $F_L$, which is
closely related to the gluon at small $x$, \emph{must} be
positive. The CT(EQ) analyses do not allow such negative gluon
  variants. We have checked for HERAPDF1.5 that the (moderately)
  negative gluon does not lead to negative $F_L$. The MSTW2008 set
  however includes member PDF sets with negative gluons that do lead to
  negative $F_L$ and are thus unphysical.

 \begin{figure}[!t]
  \vspace{5mm}
  \centering
  \includegraphics[width=\columnwidth, trim= 0cm 1.3cm 0cm 0cm, clip]{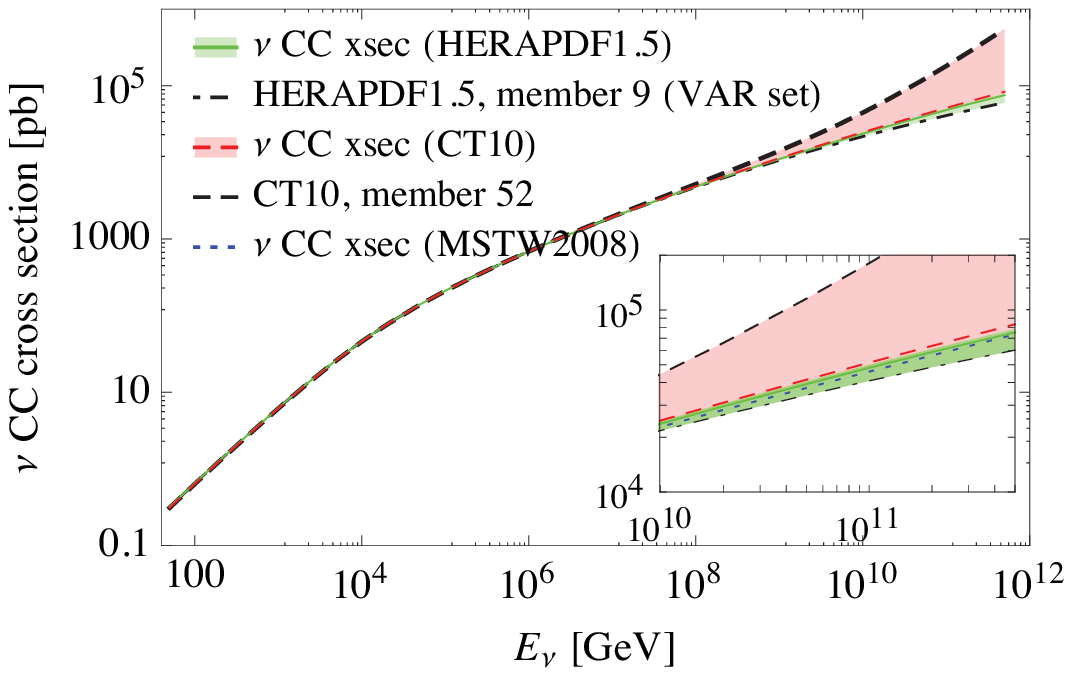}
  \includegraphics[width=\columnwidth]{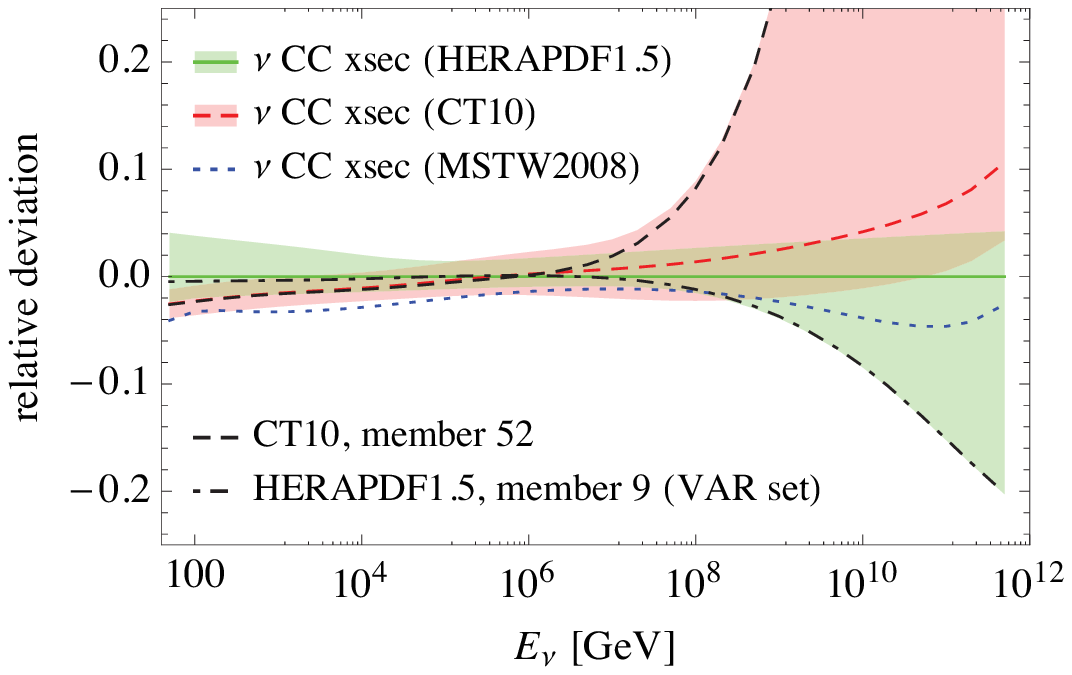}
  \caption{Comparison of the total cross-section (top panels) and
    uncertainties (bottom panels) for CC scattering as predicted by
    the HERAPDF1.5, CT10 and MSTW2008 (central member only) PDF sets.
    The cross-sections and deviations for member 9 of HERAPDF1.5 and
    member 52 of CT10 are indicated by the dashed and dot-dashed
    lines, respectively.}
  \label{fig:xsecAllWith}
 \end{figure}

 \begin{figure}[!t]
  \vspace{5mm}
  \centering
  \includegraphics[width=\columnwidth, trim= 0cm 1.3cm 0cm 0cm, clip]{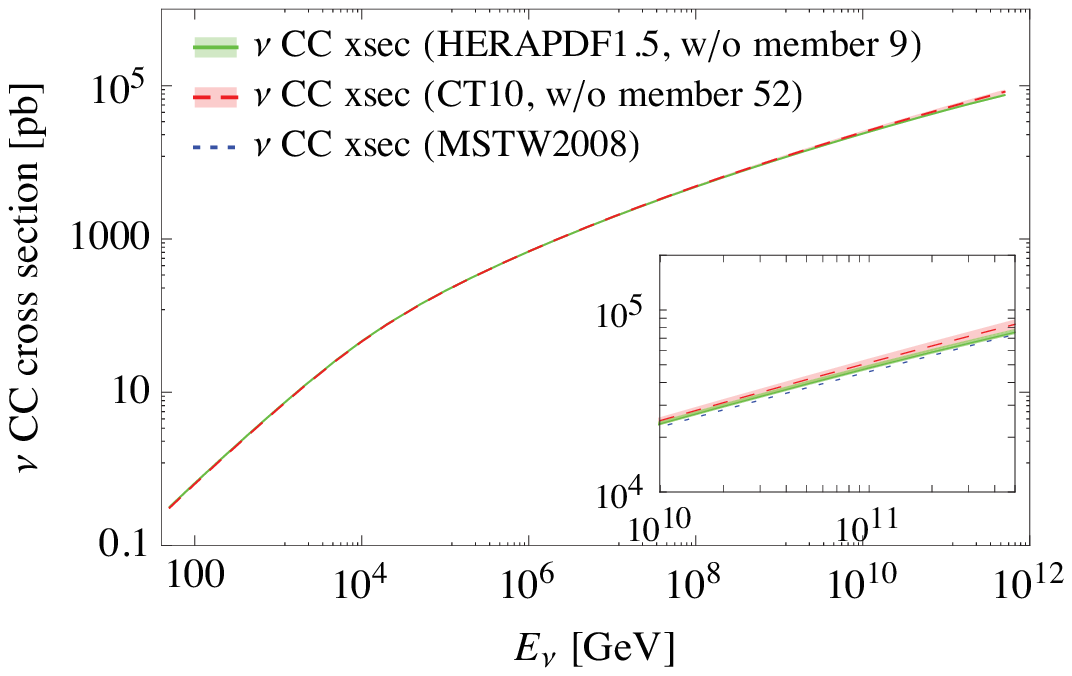}
  \includegraphics[width=\columnwidth]{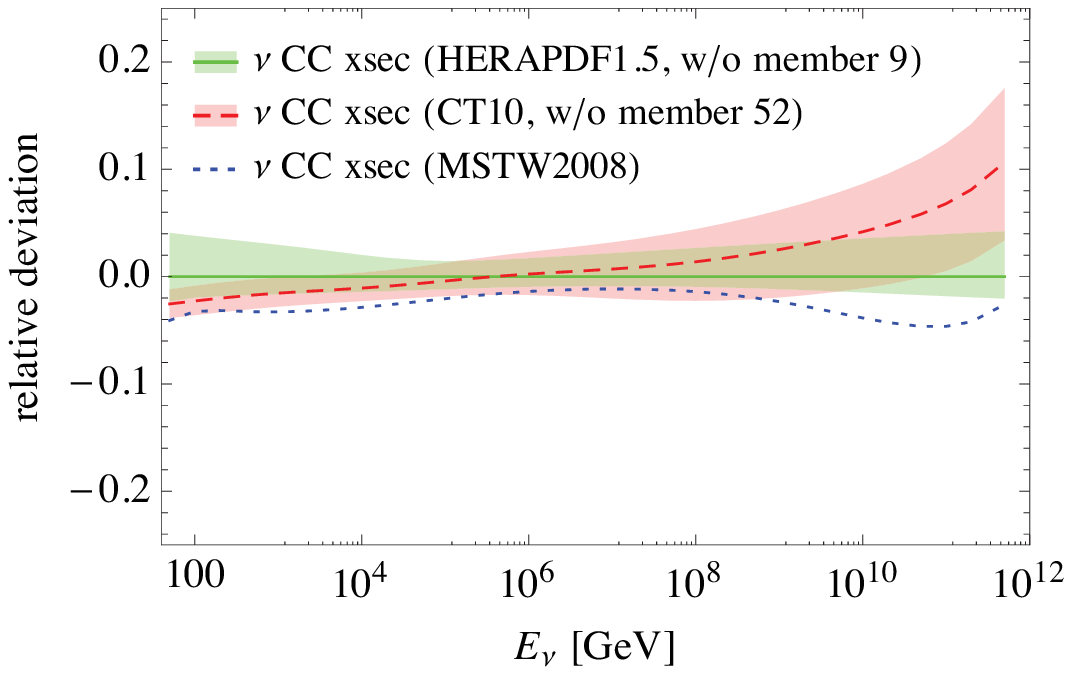}
  \caption{Same as Fig.~\ref{fig:xsecAllWith}, but excluding member 9
    of the HERAPDF1.5 set and member 52 of the CT10 set.}
  \label{fig:xsecAllWithout}
 \end{figure}

In Fig~\ref{fig:xsecAllWith} (top panel) we compare the CC
cross-sections, along with their total uncertainties (including that
coming from the variation of $\alpha_\mathrm{s}(M_Z)$), as predicted
by HERAPDF1.5 and CT10. The MSTW2008 central prediction is also
included for comparison. In Fig~\ref{fig:xsecAllWith} (bottom panel)
we emphasize the small differences in the central values of the PDFs
and their relative uncertainties. In order to highlight the effect of
the extreme members of HERAPDF1.5 and CT10 in
Figs~\ref{fig:xsecAllWithout}, we show these plots without member 9 of
the HERAPDF15 variations (which allows for the gluon to become
negative at low $x$ and $Q^2$) and without member 52 for CT10 (the
cross-section for which rises $\propto E_\nu^{0.7}$ whereas for the
central member it rises $\propto E_\nu^{0.3}$). However \textit{any}
power-law rise in the cross-section will eventually violate the
Froissart bound, which requires the rise to be no faster than $\log^2
s$~\cite{Fiore:2005wf}. This should result in a reduction of the
cross-section at high energies, by a factor of $\sim 2$ at $E_\nu =
10^{12}$~GeV~\cite{Block:2010ud} and perhaps even more
\cite{Illarionov:2011wc}.

\section{Conclusions}

We find that the predictions of high energy neutrino DIS
cross-sections from the central values of HERAPDF1.5, CT10 and
MSTW2008 PDFs are very similar. However the predictions for the
uncertainties (deriving from the uncertainties on the input PDFs)
differ quite strongly. If we exclude error sets which either lead to
too steep a rise in the cross-section, or allow the low $x$ gluon to
be negative at low $Q^2$, then we find that the uncertainty estimates
of HERAPDF1.5 and CT10 --- both of which use the most up-to-date,
accurate HERA data --- are remarkably consistent. In particular,
  we find the uncertainties to be much smaller than claimed
  recently~\cite{Connolly:2011vc}.

 \begin{figure*}[th]
  \centering
  \includegraphics[width=\textwidth]{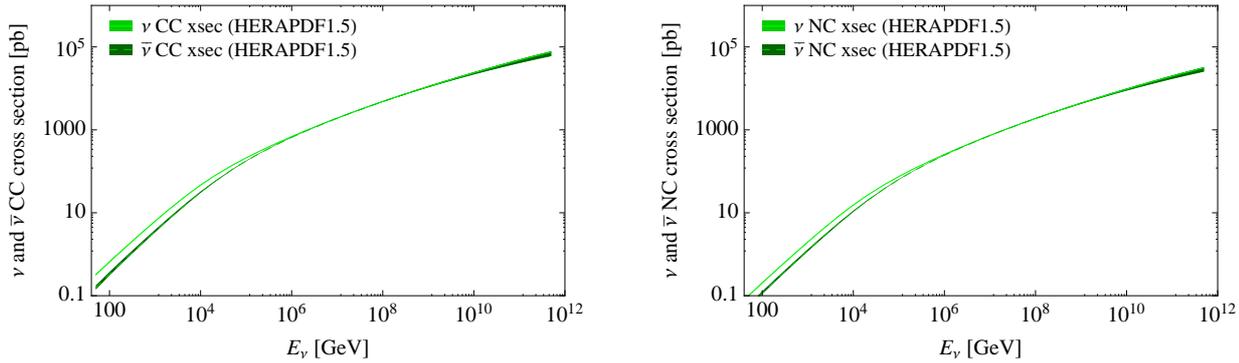}
  \caption{Neutrino and anti-neutrino cross-sections on isoscalar
  targets for CC and NC scattering for HERAPDF1.5.
    }
  \label{fig:nuANDnubarHERA}
 \end{figure*}

Our results for the high energy neutrino and anti-neutrino CC and NC
DIS cross-sections and their uncertainties using HERAPDF1.5 at NLO are
shown in Fig.~\ref{fig:nuANDnubarHERA}. The general trend of the
uncertainties can be understood by noting that as one moves to higher
neutrino energy one also moves to lower $x$ where the PDF
uncertainties are increasing. The PDF uncertainties are smallest at
$10^{-2} \lesssim x \lesssim 10^{-1}$, corresponding to $s \sim
10^5$~GeV$^2$.  When the high $x$ region becomes important the
neutrino and anti-neutrino cross-sections are different because the
valence contribution to $xF_3$ is now significant.  This is seen in
Fig.~\ref{fig:nuANDnubarHERA}, as is the onset of the linear
dependence of the cross-sections for $s < M_W^2$. Note that our
predictions are made for $Q^2 > 1$~GeV$^2$ since perturbative QCD
cannot sensibly be used at lower values.  For higher energies, we
intend to upgrade ANIS \cite{Gazizov:2004va} to use the HERAPDF1.5
(differential) cross-sections.  Meanwhile, the tabulated cross
  sections for protons, neutrons and isoscalar targets are available
  from a webpage \cite{Mandywebpage}; differential cross sections are
available upon request. Any measured deviation from these values would
signal the need for new physics beyond the DGLAP formalism.

\end{document}